\documentclass[
 amsmath,amssymb,
 aip,
reprint]{revtex4-2}

\usepackage{graphicx}
\usepackage{dcolumn}
\usepackage{bm}
\usepackage{xcolor}
\usepackage{diagbox}
\usepackage{booktabs}
\usepackage{tabularx}
\usepackage[ruled,linesnumbered]{algorithm2e}
\usepackage{multirow}
\usepackage{makecell}
\usepackage[colorlinks]{hyperref}
\hypersetup{
    colorlinks = true,
    linkcolor=blue
}
\usepackage{subfigure}

\newcommand{\tp}{\mathsf{T}}
\newcommand{\wt}[1]{\widetilde{#1}}
\newcolumntype{Y}{>{\centering\arraybackslash}X}

\newcolumntype{F}[1]{>{\centering\arraybackslash}p{#1}}

\begin{document}

\preprint{APS/123-QED}

\title{Ultra-low-energy defibrillation through adjoint optimization}
\author{Alejandro Garzon}
\email{alejandro.garzon@usa.edu.co}
\affiliation{Department of Mathematics, Universidad Sergio Arboleda, 
Bogot\'a 110221, Colombia}
\author{Roman O. Grigoriev}
\affiliation{School of Physics, Georgia Institute of Technology, Atlanta, Georgia 30332-0430, USA
}

\date{\today}

\begin{abstract}
This study investigates ultra-low-energy defibrillation protocols using a simple two-dimensional model of cardiac tissue. We find that, rather counter-intuitively, a single, properly timed, biphasic pulse can be more effective in defibrillating the tissue than low energy antitachycardia pacing (LEAP) which employs a sequence of such pulses, succeeding where the latter approach fails. Furthermore, we show that, with the help of adjoint optimization, it is possible to reduce the energy required for defibrillation even further, making it three orders of magnitude lower than that required by LEAP. Finally, we establish that this dramatic reduction is achieved through exploiting the sensitivity of the dynamics in vulnerable windows to promote annihilation of pairs of nearby phase singularities.
\end{abstract}

\maketitle

\begin{quotation}
Fibrillation is a type of cardiac arrhythmia that can be lethal unless quickly terminated. Existing protocols for terminating fibrillation employ a sequence of one or more electric pulses with a predefined temporal profile. These protocols require substantial energy and have many associated adverse effects, including intense pain and even heart tissue damage. The present study introduces an optimization approach which enables reduction of the energy required for defibrillation by several orders of magnitude by allowing the applied electric field to have an arbitrary temporal profile.    
\end{quotation}

\section{Introduction}
Fibrillation is a state of irregular and uncoordinated contraction of the cardiac muscle that can affect the ventricles or the atria, with serious physiological consequences. Ventricular fibrillation (VF), unless terminated, leads to sudden cardiac death, a major cause of mortality \cite{wong2019epidemiology}, responsible for more than 370,000 deaths in the USA during 2019 \cite{tsao2022heart}. Atrial fibrillation (AF), while not fatal, is linked to an elevated risk of detrimental outcomes, like heart failure, dementia, and stroke \cite{munger2014}. A 2010 study \cite{chugh2014worldwide} found 33.5 million individuals affected by atrial fibrillation worldwide, with 5 million new cases occurring that year, making this condition a global epidemic.

Fibrillation can be terminated by a high-energy electric shock that depolarizes a significant portion of the heart muscle, allowing restoration of the sinus rhythm. This is the standard approach used by both external defibrillators and implantable cardioverter defibrillators (ICD), proven to extend the life of patients at high risk of lethal arrhythmias \cite{passman2009shouldn}. However, these shocks tend to cause heart tissue damage \cite{babbs1980therapeutic,epstein1992gross} and intense pain, possibly leading to anxiety, depression, and post-traumatic stress \cite{sears2011posttraumatic,jacq2009comparison}.

Therefore, concerted effort has been devoted to developing low-energy defibrillation therapies. One of the oldest approaches is to use a time-periodic electric field with the frequency nearly matched to the natural frequency of the spiral waves to induce their resonant drift \cite{agladze1987} out of the tissue. As shown by Biktashev and Holden \cite{biktashev1994}, this can lead to an order of magnitude reduction in the energy, although this approach has a number of complications associated with inhomogeneity of the tissue and interaction of spiral waves with the tissue boundaries \cite{biktashev1995}.

An alternative approach, called Low Energy Antitachycardia Pacing (LEAP), uses a periodic sequence of lower energy pulses aiming to increase spatial synchronization. Several theoretical and experimental studies \cite{Luther2011,buran2017control,ji2017synchronization} have shown that LEAP can eliminate fibrillation even when the energy of each pulse is an order of magnitude lower than that of the single-shock defibrillation protocol. An improvement to LEAP has been proposed by Lilienkamp {\it et al}. \cite{lilienkamp2022taming}, who found in simulations that a gradual lengthening of the time interval between consecutive pulses, in contrast to a fixed interval, can increase the probability of defibrillation. Similarly, a modulation of the interval between pulses, aimed at 
reducing the intercellular dispersion of so-called isostable coordinates, was proposed by Wilson and Moehlis \cite{wilson2016toward}. A recent clinical investigation in humans \cite{ng2021novel} verified successful termination of atrial fibrillation by means of a sequence of pulses of varying energy, some of those possibly below the pain threshold.  DeTal {\it et al.} \cite{detal2022terminating} demonstrated a procedure to identify tissue regions where application of a single low-energy pulse by multiple electrodes can terminate reentrant waves.

These studies suggest that painless defibrillation via a sequence of low-energy pulses might become a reality in clinical practice, in the near future. However, even if this favorable scenario materializes, the quest for a further reduction of the energy of the defibrillating protocols should not stop, as pain is only one adverse effect. It is not unreasonable to suspect long-term tissue damage even from low-energy pulses considering that the termination of AF might be, for some patients, a frequent occurrence. For instance, the investigation of Gold {\it et al.} \cite{gold2001clinical} on the efficacy of an ICD to stop AF found a mean rate of seven treated episodes of atrial tachyarrhythmias per patient-month. Moreover, these numerous treatments imply fast battery drainage, reducing the lifetime of an ICD and increasing mortality risks associated with its surgical replacement \cite{merchant2016implantable}. In this regard, it is worth noting that, for some therapies based on sequences of low energy pulses, although the per-pulse energy is lower, compared to single shock therapy, the total energy is similar \cite{ji2017synchronization}, so LEAP might not increase the lifetime of ICDs.

Given the potential advantages of a further reduction of the energy required for defibrillation, the question arises as to both how defibrillation protocols with an even lower energy can be identified and how much the energy can be reduced. Single-shock defibrillation offers little room for improvement, although some signal shape optimization has been attempted \cite{bragard2013shock,chamakuri2015application}.
LEAP offers more opportunities for optimization by varying the number of pulses, the inter-pulse time, and the single-pulse energy; searches in this low-dimensional parameter space
have been attempted as well \cite{buran2017control}. This approach has the limitation of restricting the search to a narrow family of signals. 

In the present work, we describe a systematic procedure for finding defibrillating signals with low energy in a two-dimensional model of cardiac tissue representing the atria. The search space is expanded to include continuous functions of time, representing an externally applied electric field, on a finite time interval. Our approach aims to minimize the total energy of the defibrillation protocol while simultaneously maximizing the success rate using adjoint optimization. A similar approach has been used previously in a much more restricted capacity to optimize the signal shape for single-shock defibrillation on a very brief temporal interval (4 ms) \cite{chamakuri2015application}. In contrast, in this work, we consider signals over much longer time intervals comparable to the LEAP protocol, i.e., several characteristic periods of spiral wave revolution.

The paper is structured as follows. In Sec. \ref{sec:model}, the electrophysiological model, the optimization problem, and its solution through adjoint optimization are described. The results are presented in Sec. \ref{sec:spiral} which compares different defibrillation protocols. Conclusions and perspectives are offered in Sec. \ref{sec:conclusions}.

\section{Problem description}
\label{sec:model}

To highlight the generality of our approach, it is discussed in terms of an electrophysiological model of cardiac tissue with $n$ state variables that can describe atria (or ventricles) when defined in two (or three) spatial dimensions. Let $u_1(t,{\bf r})$ represent the transmembrane voltage and $u_2(t,{\bf r}),\cdots,u_n(t,{\bf r})$ be the gating variables. Here $t$ denotes time and ${\bf r}$ denotes the spatial position of cells in the domain $\Omega$, which contains non-conducting patches representing anatomical heterogeneities, such as blood vessels. The dynamics are described by a system of coupled partial differential equations
\begin{align}\label{eq:puwtLu}
    \partial_t{\bf u}&=\widetilde{L}{\bf u} + \wt{F}({\bf u}),
\end{align}
where ${\bf u}=[u_1,u_2,...,u_n]^\tp$
is the state vector, and
$\widetilde{L}{\bf u}=[D_1 \nabla^2u_1,D_2 \nabla^2u_2,...,D_n \nabla^2u_n]^\tp$ describes the electrical coupling between cells. A small diffusion coefficient $D_i \ll D_1,~i=2,...,n$, is assumed for the gating variables, in order for the model to possess smooth solutions. Finally, $\wt{F}({\bf u})$ represents the local membrane dynamics (ionic model). 

In the presence of an external electric field ${\bf E}(t,{\bf r})$, the scaled voltage $u_1$ satisfies the boundary condition \cite{pumir1999}
\begin{align}
  \label{eq:nu1E}
    {\bf n}\cdot(\nabla u_1 - {\bf E})= 0,
\end{align}
where ${\bf n}$ is the unit vector normal to the domain boundary $\Gamma=\partial\Omega$, which includes the boundaries of the heterogeneities. For the gating variables $u_i$, $i=2,...,n$, no-flux boundary conditions are assumed:
\begin{align}
  \label{eq:nui}
      {\bf n}\cdot\nabla u_i= 0.
    \end{align}
In the absence of an external electric field, the system has (i) a uniform equilibrium solution ${\bf u}(t,{\bf r}) = {\bf u}^*$ which describes the (stable) quiescent state corresponding to a total lack of electrical activity in the tissue and (ii) solutions describing persistent spiral wave chaos which represents fibrillation. The goal of defibrillation is therefore to suppress chaotic dynamics in favor of the quiescent state by means of an appropriately chosen external electric field ${\bf E}(t,{\bf r})$.
    
For simplicity, we assume the applied electric field to be spatially uniform but varying in time. Choosing the $x$-axis pointing in the direction of the field, we have ${\bf E}(t,{\bf r})=E(t){\bf\hat{x}}$, where the hat denotes a unit vector. A time-dependent electric field $E(t)$ applied within a finite temporal window $0<t<T$  leads to successful defibrillation if ${\bf u}(t,{\bf r})$ approaches the quiescent state ${\bf u}^*$ within this finite window or soon after. Up to a dimensional prefactor, the total energy delivered to the tissue is given by
    \begin{align}
     \mathcal{N} = \int_0^T [E(t)]^2dt.
   \end{align}
The quiescent state ${\bf u}^*$ is spatially uniform, but it does not need to be reached by $t=T$. In fact, defibrillation can be achieved by simply eliminating spatial gradients (i.e., dispersion) of all the state variables by this time. Hence an optimal defibrillating signal can be found by minimizing the functional $\mathcal{L}$
\begin{align}
  \label{eq:L0}
    \mathcal{L}=\frac{\alpha}{2}\mathcal{N}+\frac{1}{2}\mathcal{M},
   \end{align}
where
\begin{align}
  \label{eq:calJi}
  \qquad
    \mathcal{M}=\sum_{i=1}^n\gamma_i \mathcal{J}_i,\quad \mathcal{J}_i = \int_\Omega |\nabla u_i|_{t=T}^2\,d\Omega,
\end{align}
and  $\alpha$ and $\gamma_i$ are positive parameters.
The first term in \eqref{eq:L0} represents the total energy and the second term minimizes the gradients of the voltage and gating variables at the end of the interval.
Parameters $\alpha$ and $\gamma_i$ can be tuned to emphasize the relative importance of the gradients and the energy of the defibrillation signal. For instance, large values of $\gamma_i$'s will result in a faster but less energy-efficient defibrillation protocol while smaller values will result in a slower but more energy-efficient defibrillation protocol.

For fixed model parameters, initial condition ${\bf u}_0({\bf r})$, and the final time $T$, $\mathcal{L}$ depends solely on $E(t)$. In other words, it is a functional defined over the set of all possible $E(t)$. To minimize $\mathcal{L}$, the gradient descent method will be used, which requires the calculation of the associated functional derivative
\begin{align}
\mathcal{G}(t)=\frac{\delta\mathcal{L}}{\delta E(t)},\quad 0 \le t \le T.
\end{align}
The procedure used to evaluate this derivative is presented in Appendix \ref{sec:fin_diff}. 
Let us consider a continuous family of functions $E_s(t)$, where $s$ is a continuous parameter, such that 
\begin{align}\label{eq:dEds}
  \frac{dE_s(t)}{ds}=-\mathcal{G}(t)|_{E_s(t)}.
\end{align}
Then $\mathcal{L}[E_s(t)]$ is a non-increasing function of $s$ with the minima corresponding to a vanishing $\mathcal{G}(t)$. An optimal electric field minimizing $\mathcal{L}$ can therefore be found by solving the differential equation \eqref{eq:dEds} numerically, e.g, using finite differences, starting from an appropriate initial condition $E_0(t)$. 

The simplest way to compute the functional derivative is using finite differences, as explained in Appendix \ref{sec:fin_diff}. For a numerical solution evaluated on a discrete temporal grid $t_l = l\Delta t,~l=0,1,..., N$, where $\Delta t=T/N$ is the time step, the cost of evaluating $\mathcal{G}(t)$ scales as $N^2$. This evaluation becomes prohibitively expensive for longer temporal intervals. We therefore employed an alternative method for calculating $\mathcal{G}(t)$ whose cost scales linearly with $N$. This approach, known as the adjoint method, is more clearly explained in terms of the spatial discretization of the model \eqref{eq:puwtLu}, as described next.

\subsection{Spatial discretization}

The spatial domain $\Omega$ was chosen as a square containing circular (non-conducting) heterogeneities with radii and locations selected randomly. The radii followed the power-law probability distribution \cite{Luther2011}, whereas the locations were distributed uniformly. Enough locations were selected as to yield an average density of 16 heterogeneities per ${\rm cm}^2$. The solution ${\bf u}(t,{\bf r})$ of \eqref{eq:puwtLu} was evaluated on a uniform square grid with 256 points on each side, with the grid points inside the heterogeneities excluded. This resulted in a computational grid with $m=64,071$ points. The grid spacing was chosen to be $\Delta x = \Delta y = 0.035$ cm, which translates to a domain side length of $8.925$ cm.

Let ${\bf w}$ denote the discretization of ${\bf u}(t,{\bf r})$, where ${\bf w} = [u_{11},\cdots, u_{1m},\cdots,u_{n1},\cdots,u_{nm}]^\tp$, $u_{ij} \approx u_i({\bf r}_j)$, and ${\bf r}_j$ is the position of the $j$-th grid point.  Second-order finite differences were used to represent the Laplacians $\nabla^2 u_i$ and the derivatives in the boundary conditions \eqref{eq:nu1E} and \eqref{eq:nui}. The dynamics of the tissue is then described by a system of coupled ODEs 
\begin{subequations}
\label{eq:main_ivp_bfu}
\begin{align}
  \label{eq:main_dot_bfu}
  \dot{\bf w} &= L{\bf w} + F({\bf w}) + E{\bf b},\\
  \label{eq:main_dot_bfuT}
  {\bf w}(0) &= {\bf w}_0,
\end{align}
\end{subequations}
where $L$ and ${\bf F}$ represent discretizations of $\wt{L}$ and $\wt{\bf F}$, respectively,
and the term $E(t){\bf b}$ represents the effect of the electric field.
The cost function \eqref{eq:L0}
can also be expressed in terms of ${\bf w}$:
\begin{align}
  \label{eq:calM}
  \mathcal{M}
  \approx  \left[{\bf w}^\tp R {\bf w}\right]_{t=T},
\end{align}
where $R$ is 
a block-diagonal matrix.
The details are given in Appendix \ref{sec:discretization}.

\subsection{Adjoint method}
\label{sec:adjoint}
We will follow the ``adjoint looping'' approach used previously in geophysics \cite{plessix2006review}, fluid dynamics \cite{pringle2010using,krause2013adjoint}, and plasma physics \cite{nies2022adjoint}. It involves the adjoint field ${\bm{\lambda}}(t)$ which is a solution of the initial value problem
\begin{subequations}
  \label{eq:main_ivp_lamb}
  \begin{align}
    \label{eq:main_dot_lamb}
    -\dot{\bm{\lambda}} &= (L+J_F)^\tp \bm{\lambda}, \\
    \label{eq:main_lambT}
    \bm{\lambda}(T) &= R{\bf w}(T),
  \end{align}
\end{subequations}
where $J_F={dF}/d{\bf w}$ is the Jacobian of $F({\bf w})$. The derivation is presented in Appendix \ref{sec:adj_method}.

The adjoint method involves an iterative procedure entailing the following steps. First, the boundary value problem \eqref{eq:main_ivp_bfu} is solved forward in time, which determines ${\bm\lambda}(T)$. Next, $\bm{\lambda}(t)$ is found by solving the initial value problem \eqref{eq:main_ivp_lamb} backward in time. Note that solution of this problem requires evaluation of the Jacobian $J_F$ at each time  $0\le t\le T$, which requires the entire solution ${\bf w}(t)$ to be stored. Next, the functional derivative of the cost function is evaluated  as 
\begin{align}
   \mathcal{G}(t) = \alpha E(t) + {\bf b}^\tp\bm{\lambda}(t).
\end{align}
All variables are evaluated using $E(t)=E_s(t)$.
Finally, the electric field is updated according to 
\begin{align}
  \label{eq:grad_desc}
  E_{s+\Delta s}(t) = E_s(t) - \Delta s\, \mathcal{G}(t)|_{E_s(t)}
  \end{align}
with a sufficiently small $\Delta s$.
The entire process is then repeated until convergence of $E(t)$ is achieved.

\section{Spiral wave chaos and defibrillation protocols}
\label{sec:spiral}

In this work, simulations were carried out for the three-variable Fenton-Karma model \cite{Fenton1998} of ionic dynamics. The parameters used are shown in Table \ref{tab:param}. To facilitate computation of spatial derivatives of the gating variables and the Jacobian $J_F$, the Heaviside step functions $\Theta(\chi)$  used in the model were replaced by their smoothed analogues
\begin{align}
  \label{eq:Theta_s}
\Theta_a(\chi) = \frac{1+ \tanh(\chi/a)}{2},
\end{align}
with parameter $a=0.01$ defining the width of the ``step.'' The diffusion coefficients were set to $D_1=10^{-3}\;{\rm cm}^2/{\rm ms}$ and $D_2=D_3=10^{-5}\;{\rm cm}^2/{\rm ms}$. Solutions of Equations \eqref{eq:main_ivp_bfu} and \eqref{eq:main_ivp_lamb} were computed using fourth-order Runge-Kutta method with a time-step size $\Delta t = 0.1\;{\rm ms}$. To speed up the computations, the integration function was implemented using CUDA (Nvidia Inc.) for execution on general-purpose graphics processing units. For ease of use, the CUDA code was then wrapped in MATLAB (Mathworks Inc.) mex-functions.

A pair of reentrant waves was generated using an S1-S2 protocol. First, the quiescent state $(u_1,u_2,u_3)=(0,1,1)$ was excited at a corner (S1 stimulus) by setting $u_1=1$ to produce a circular excitation wave. Note that the transmembrane voltage is given by $V = V_0 + (V_{\rm fi}-V_0)u_1$, where $V_0 = -85\;{\rm mV}$ is the reference voltage corresponding to the quiescent state and $V_{\rm fi} = 15\;{\rm mV}$\cite{Fenton1998}. A short time after the waveback had passed the center of the domain, a small patch of tissue was excited at the center (S2 stimulus). From the stimulated patch, a circular wave develops and produces a conduction block when meeting the back of the previous wave, generating a pair of spiral waves.

\begin{figure*} 
  \centering
  \includegraphics[width=\columnwidth]{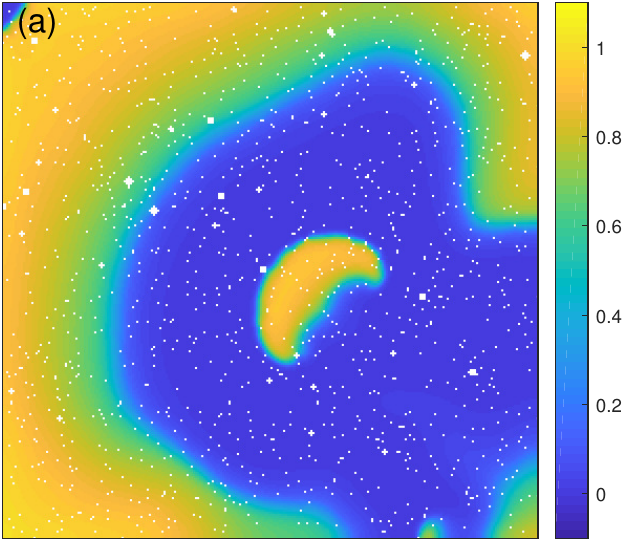}
  \includegraphics[width=\columnwidth]{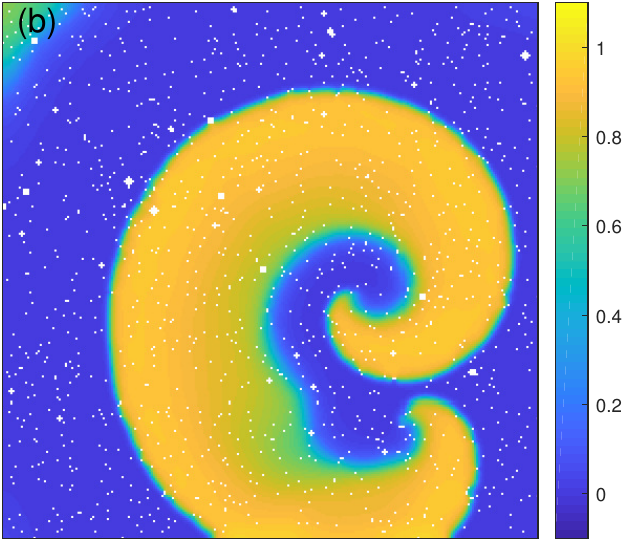}\\ \vspace{3mm}
  \includegraphics[width=\columnwidth]{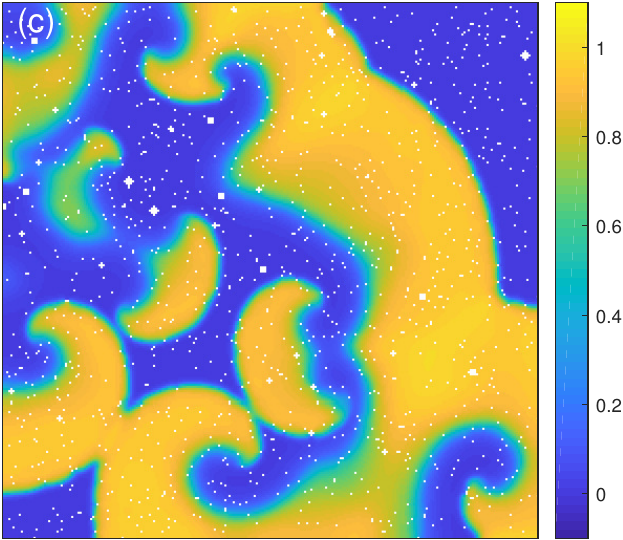}
  \includegraphics[width=\columnwidth]{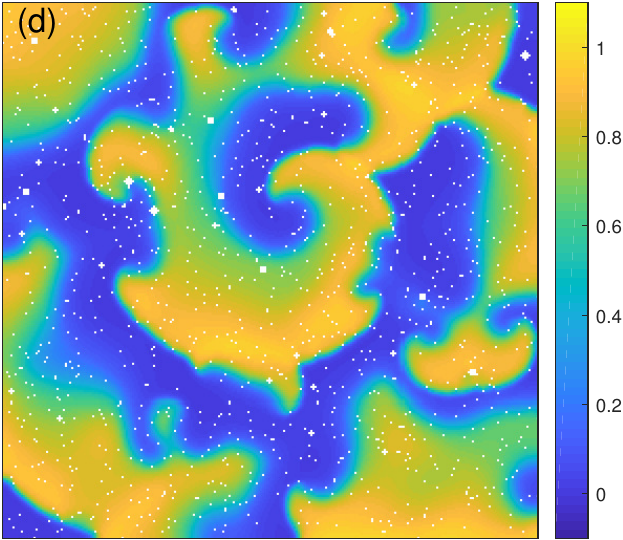}\\ \vspace{3mm}
  \includegraphics[width=\columnwidth]{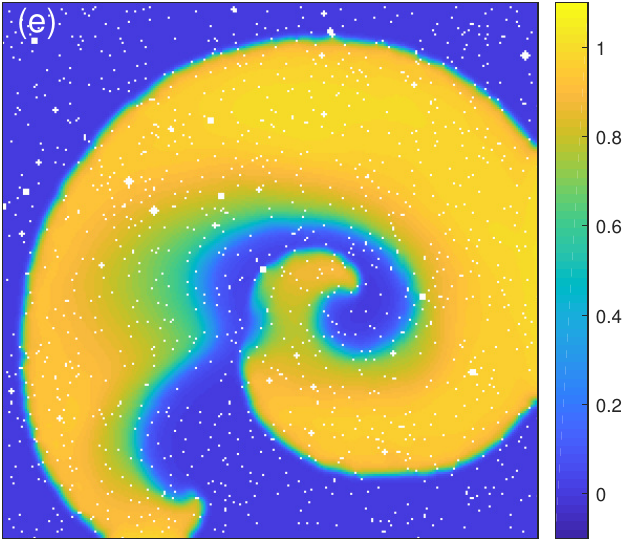}
  \includegraphics[width=\columnwidth]{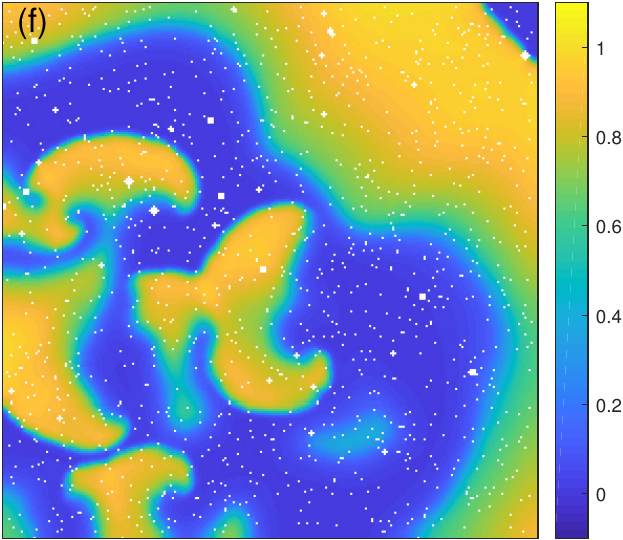}
  \caption{(Multimedia available online) Snapshots of the voltage field ($u_1$) describing a state of fibrillation at times (a) $t=0$ ms, (b) $t=290$ ms, (c) $t=2.09$ s, (d) $t=6.83$ s, (e) $t=13.34$ s, and (f) $t=20$ s. Nonconducting heterogeneities are shown in white.}
  \label{fig:snap}
\end{figure*}

Figure \ref{fig:snap}(a)  shows the voltage ($u_1$) about 1 s after the S2 stimulus. In the absence of electric field $E(t)$, this initial condition gives rise to a state of persistent spiral wave chaos that lasts more than 20 s. For reference, the characteristic revolution time for a spiral wave is two orders of magnitude smaller, as shown below. Hence, in this system, spiral wave chaos represents either a sustained state (an attractor) or a long-lived transient (chaotic repeller). Figure \ref{fig:snap} displays characteristic snapshots of the voltage field at different times with a corresponding movie provided as a multimedia file (available online). The state shown in Figure \ref{fig:snap}(a) was used as a representative initial condition ${\bf u}_0$ to compare the different defibrillation protocols discussed below. Other choices, such as the states shown in Figures \ref{fig:snap}(b-e) produced qualitatively similar results and hence are not discussed here.

\begin{table}[h]
  \centering
  \renewcommand{\arraystretch}{1.2}
  \begin{tabular}{cD{.}{.}{3.3}cD{.}{.}{3.3}cD{.}{.}{3.3}}
    \hline\hline
    Parameter & \multicolumn{1}{c}{Value} & Parameter & \multicolumn{1}{c}{Value} & Parameter & \multicolumn{1}{c}{Value} \\ \hline
    $u_c$         & 0.13 &    $\tau_v^+$ & 3.324 &  $k$ & 15 \\           
    $u_v$         & 0.04 &    $\tau_0$ & 8.2    & $\tau_w^-$ & 68 \\ 
    $\tau_d$      & 0.388 &   $\tau_r$ & 33.264 &  $\tau_w^+$ & 400 \\          
    $\tau_{v1}^-$ & 7.6   & $\tau_{si}$ & 29.0     &     &   \\         
    $\tau_{v2}^-$ & 8.8   & $u_c^{si}$ & 0.54    &    &    \\      \hline\hline          
  \end{tabular}
  \caption{Parameters used in the Fenton-Karma model. All time scales are in units of milliseconds. The rest of the parameters are nondimensional.}
  \label{tab:param}
\end{table}

\subsection{Single-pulse defibrillation}
\label{sec:single}
\begin{figure} 
  \centering  \includegraphics[width=\columnwidth]{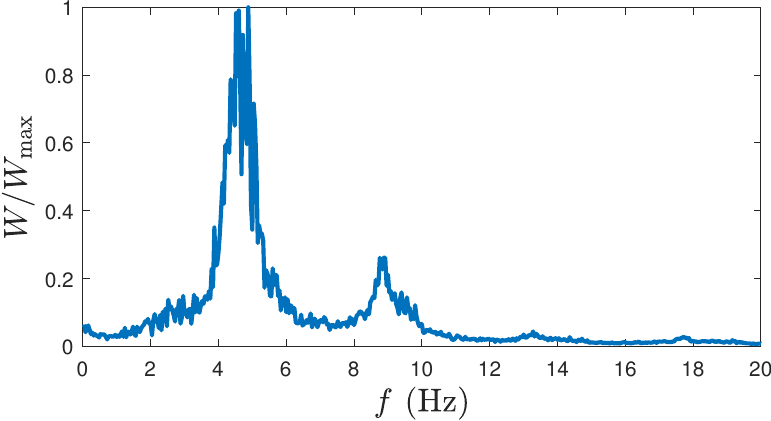}
  \caption{Power spectrum of the electrical activity normalized by $W_{\rm max}=\max_f W(f)$.}
  \label{fig:fourier}
\end{figure}

Conventional biphasic single-pulse defibrillation protocol and LEAP were explored to obtain a set of reference results. A key parameter of LEAP is the typical period of the electrical activity during fibrillation, which corresponds to one revolution of a spiral wave. That period can be identified with the help of the power spectral density $W(f)$ (see Appendix \ref{sec:spectral}). A sample power spectrum for the solution with the initial condition depicted in Figure \ref{fig:snap}(a) is shown in Figure \ref{fig:fourier}.  The maximum of $W(f)$ at around $f=4.7\;{\rm Hz}$ corresponds to a typical period $P\approx 210\;{\rm ms}$.

To examine single-pulse defibrillation, an asymmetric biphasic pulse was applied to the tissue,
\begin{align}
  \label{eq:single}
  E(t) = E^*\,g_p(t-t^*),
\end{align}
where $E^*$ is the pulse strength, $t^*$ is the pulse timing and
\begin{align}
  \label{eq:gp}
  g_p(t) = \Theta_a(t) - 2\Theta_a\!\left(t-0.7p\right) + \Theta_a(t-p)
\end{align}
with $\Theta_a$ defined in Equation \eqref{eq:Theta_s}, $a=0.1\;{\rm ms}$, and the pulse duration is $p=10\;{\rm ms}$. Pulses with smooth edges were used to allow the application of high-order integration methods for Equation \eqref{eq:main_dot_bfu}. The pulse timing was changed from $t^*=0$ ms up to $t^*=200\;{\rm ms}$ in steps of $20\;{\rm ms}$, to target the different phases of the excitation dynamics. 
For simplicity, we consider fibrillation terminated when a state with transmembrane voltage $u_1 < 0.1$ in the entire tissue domain is reached (it is straightforward to reformulate this condition in terms of the spatial heterogeneity $\mathcal{M}$). An instance of successful defibrillation was declared if such a state was reached during an interval of time less than $5P=1050\;{\rm ms}$, measured from the end of the pulse. Table \ref{tab:single} shows our results for different pulse strengths $E^*$  and the associated energies $\mathcal{N}$. Note how the likelihood of defibrillation decreases as the pulse strength $E^*$ diminishes, so that for $E^*=0.5\;{\rm V}/{\rm cm}$ no defibrillation is observed for the values of $t^*$ examined.

\begin{table*}[t]
  \centering
  {\renewcommand{\arraystretch}{1}
    \setlength{\extrarowheight}{1pt}

  \begin{tabular}{D{.}{.}{-1}F{2.6cm}*{11}{F{0.8cm}}}
    \toprule
    \multicolumn{1}{F{1.2cm}}{\multirow{2}{=}{ \makecell{$E^*$\\$({\rm V}/{\rm cm})$} }} & 
    \multirow{2}{=}{ \makecell{$\mathcal{N}$\\$({\rm V}^2{\rm ms}/{\rm cm}^2)$} } &
    \multicolumn{11}{c}{$t^*$ (ms)} \\ \cline{3-13}
       &   & 0 & 20 & 40 & 60 & 80 & 100 & 120 & 140 & 160 & 180 & 200 \\
    \midrule
    0.5 & $2.43$ &      $\square$ &      $\square$ &      $\square$ &      $\square$ &      $\square$ &      $\square$ &      $\square$ &      $\square$ &      $\square$ &      $\square$ &      $\square$ \\
    1   & $9.72$ &      $\square$ & $\blacksquare$ & $\blacksquare$ &      $\square$ &      $\square$ &      $\square$ &      $\square$ &      $\square$ &      $\square$ &      $\square$ &      $\square$ \\
    1.5 & $21.9$ &      $\square$ &      $\square$ &      $\square$ &      $\square$ & $\blacksquare$ &      $\square$ &      $\square$ &      $\square$ &      $\square$ &      $\square$ &      $\square$ \\
    3   & $87.5$ &      $\square$ &      $\square$ &      $\square$ &      $\square$ &      $\square$ & $\blacksquare$ & $\blacksquare$ & $\blacksquare$ &      $\square$ &      $\square$ &      $\square$ \\
    5   & $243$ &      $\square$ &      $\square$ &      $\square$ & $\blacksquare$ &      $\square$ & $\blacksquare$ & $\blacksquare$ & $\blacksquare$ &      $\square$ &      $\square$ &      $\square$ \\
    8   & $622$ & $\blacksquare$ &      $\square$ & $\blacksquare$ & $\blacksquare$ &      $\square$ & $\blacksquare$ & $\blacksquare$ & $\blacksquare$ & $\blacksquare$ &      $\square$ &      $\square$ \\
    \bottomrule
  \end{tabular}}
  \caption{Defibrillation by a single pulse as defined by Equation \eqref{eq:single}. Conventions: $\blacksquare$ (defibrillation), $\square$ (no defibrillation).}
  \label{tab:single}
\end{table*}

\subsection{Low-Energy Antitachycardia Pacing}
\label{sec:leap}
The likelihood of defibrillation can be increased using the LEAP protocol. This method employs a sequence of $M$ pulses, with pulse separation usually chosen close to the typical period $P$ of the fibrillatory electrical activity. For our investigation of LEAP, we used the external electric field
\begin{align}
  \label{eq:leap}
  E(t) = E^* \sum_{m=0}^{M-1} g_p[t - (t^* + mP)],
\end{align}
where the pulses are biphasic ($g_p$ is defined by Equation \eqref{eq:gp}), have the same strength $E^*$, and are applied at times $t^*, t^* + P, ..., t^* + MP$, with $P=210\;{\rm ms}$.

\begin{figure} 
  \centering
  \includegraphics[width=\columnwidth]{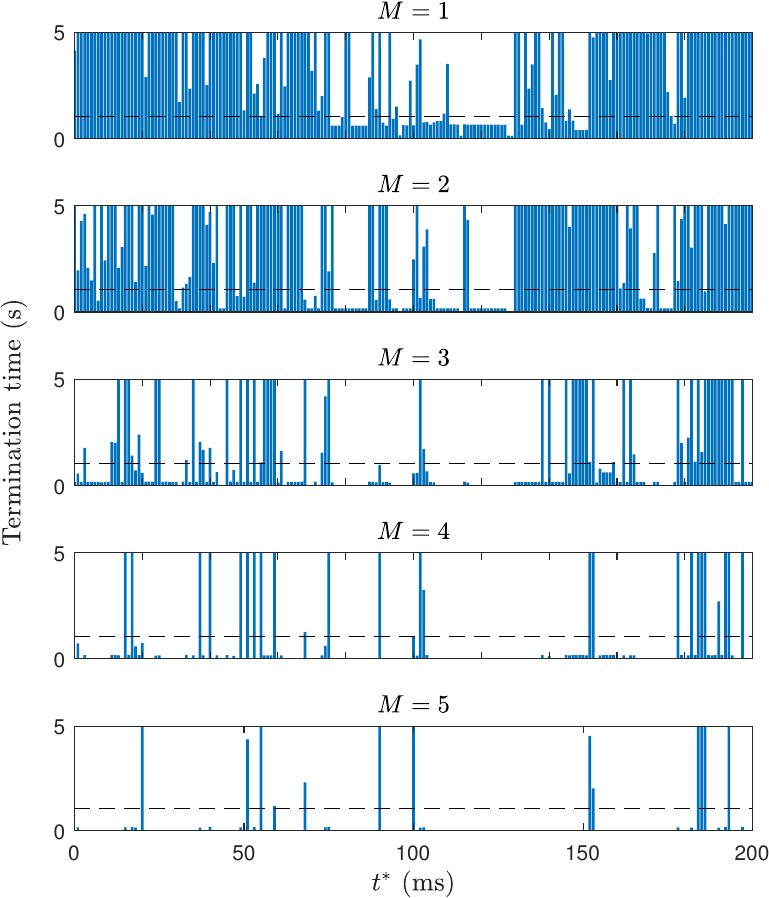}
  \caption{Termination time (bar height) of electrical activity for LEAP signals defined by Equation \eqref{eq:leap}, with  pulse strength $E^*=3\;{\rm V}/{\rm cm}$, number of pulses $M$ indicated above each panel, and timing $t^* $ varied on steps of $1\;{\rm ms}$. A bar that reaches the top edge of the axis box corresponds to a termination time above $5\;{\rm s}$, not determined, possibly infinite. On each panel, the horizontal dashed line marks the threshold of $5P$ used to declare defibrillation.}
  \label{fig:leap_E0_3}
\end{figure}

Figure \ref{fig:leap_E0_3} shows the termination time (measured from the end of the last pulse) for a pulse strength $E^*=3\;{\rm V}/{\rm cm}$, different number of pulses $M$, and time $t^*$ varied in steps of $1\;{\rm ms}$. The dashed line marks the $5P$ threshold used to declare defibrillation. Note how the likelihood of defibrillation greatly increases with the number of pulses. In Ref. \onlinecite{ji2017synchronization}, synchronization of the excitation was identified as the mechanism for LEAP defibrillation. More exactly, it was found that each LEAP pulse increases the spatial homogeneity of the phase of the excitation wave across the tissue, thus raising the probability of defibrillation. The results shown in Figure \ref{fig:leap_E0_3} are consistent with this mechanism. We also find that, even at this relatively high value of $E^*$, LEAP can fail for some choices of $t^*$. Moreover, additional pulses can have an adverse effect, with $M+1$ pulses failing to defibrillate the tissue when $M$ pulses would eventually succeed, as can be seen by comparing the results for $M=4$ and $M=5$. Nonetheless, the likelihood of successful termination of fibrillation is over 90\% for this choice of $E^*$; we can use the corresponding energy $\mathcal{N}_5=87.5\;{\rm V}^2{\rm ms}/{\rm cm}^2$ as a benchmark to compare our results against.

\begin{figure} 
  \centering
  \includegraphics[width=\columnwidth]{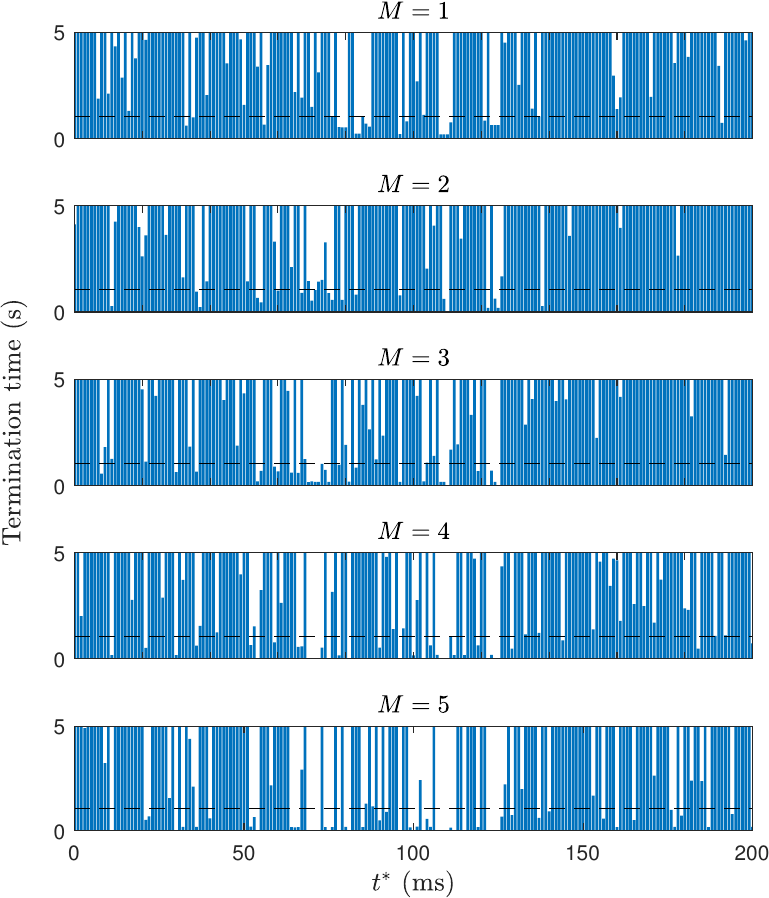}
  \caption{Termination time of electrical activity for LEAP signals with  pulse strength $E^*=1.5\;{\rm V}/{\rm cm}$, number of pulses $M$ indicated above each panel, and timing $t^* $ varied on steps of $1\;{\rm ms}$. On each panel, the horizontal dashed line marks the threshold of $5P$ used to declare defibrillation.}
  \label{fig:leap_E0_1p5}
\end{figure}

LEAP becomes progressively more inefficient in inducing synchronization as the pulse strength is decreased. Figure \ref{fig:leap_E0_1p5}, shows the termination time for $E^*=1.5\;{\rm V}/{\rm cm}$ and the same combination of values of $M$ and $t^*$ as in Figure \ref{fig:leap_E0_3}. While the likelihood of defibrillation is again found to increase with $M$, it increases much slower, as comparison of Figures \ref{fig:leap_E0_3} and \ref{fig:leap_E0_1p5} makes quite clear.

On the other hand, both Figures \ref{fig:leap_E0_3} and \ref{fig:leap_E0_1p5} have an interesting common feature: for all values of $E^*$ and $M$ considered, one finds instances of successful defibrillation flanked on both sides by failures. Similarly, one finds failures flanked by successful outcomes. The same result would be found if the resolution of sampling in terms of $t^*$ were decreased. This suggests that the measure, in terms of $t^*$, on which LEAP is successful is fractal. Indeed, a sequence of pulses can be thought of as setting up a particular initial condition ${\bf u}$ that, for a fixed $M$, is a continuous function of $E^*$ and $t^*$ in the infinite-dimensional state space of the autonomous system \eqref{eq:puwtLu}. Fractal dependence on $t^*$ suggests that the boundary between the two attractors, one representing continuing fibrillation (e.g., failed defibrillation) and the other representing the stable equilibrium (e.g., successful defibrillation) is fractal. This is analogous to what is found in transitional fluid turbulence \cite{schneider2007}, where the two attractors represent turbulent and laminar flow. It is important to note that the qualitative picture does not change even when chaotic dynamics represent a long-lived transient described by a chaotic repeller rather than a chaotic attractor.

\begin{figure} 
  \centering
  \includegraphics[width=\columnwidth]{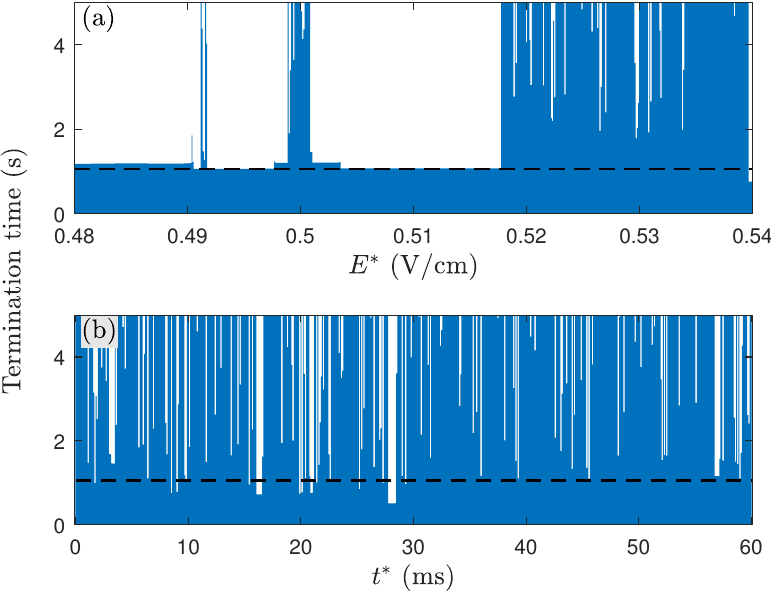}
  \caption{Termination time of electrical activity for a single low-energy pulse. (a) Fixed time $t^* = 20\;{\rm ms}$ and pulse strength $E^*$ changed on steps of $10^{-4}\;{\rm V}/{\rm cm}$. (b) Fixed $E^*=0.5\;{\rm V}/{\rm cm}$ and pulse timing $t^*$ varied on steps of $\Delta t=0.1\;{\rm ms}$. On each panel, the horizontal dashed line marks the threshold of $5P$ used to declare defibrillation.}
  \label{fig:E0_t0_t_term}
\end{figure}

To confirm the fractal structure of the basin boundary, we explored the dependence of the outcome on parameters for the single-pulse defibrillation protocol ($M=1$). Specifically, we performed two sets of calculations by varying either the pulse strength or the pulse timing. In the first case, the pulse timing was fixed at $t^* = 20\;{\rm ms}$ and the pulse strength $E^*$ was varied in steps of $10^{-4}\;{\rm V}/{\rm cm}$, around $E^*=0.5\;{\rm V}/{\rm cm}$. In the second case, the pulse strength was fixed at $E^*=0.5\;{\rm V}/{\rm cm}$ (which corresponds to the energy of $\mathcal{N}_1 = 2.43\;{\rm V}^2{\rm ms}/{\rm cm}^2$) and the pulse timing $t^*$ was varied in steps of $0.1\;{\rm ms}$. As Figure \ref{fig:E0_t0_t_term} illustrates, in both cases, we find the pattern noted previously, i.e., a sensitive, nonmonotonic, and non-smooth dependence of the defibrillation time on parameters, which is consistent with a fractal structure of the basin boundary.

These results also illustrate that it is, in principle, possible to terminate fibrillation using a very simple low-energy protocol, i.e., a properly timed {\it single} biphasic pulse, with the energy which is one or two orders of magnitude smaller than that of a typical LEAP sequence. While the requirement to finely tune the parameters (here, $E^*$ and $t^*$) makes this approach impractical, its success raises the fundamental question of whether the energy can be reduced even further by optimizing the temporal profile of the electric field $E(t)$. Hence, in the next section, we use the adjoint method to construct such optimal profiles.

\subsection{Optimal defibrillation protocols}
\label{sec:optimal_defib}

The functional $\mathcal{L}$ is defined over a space of functions $E(t)$ with finite support $t\in [0,T]$ and, for a given initial state ${\bf u}_0$ (or its discretization ${\bf w}_0$), it can have infinitely many local minima. The distance between these minima will generally decrease as $T$ is increased, creating a rough ``landscape'' featuring both shallow and deep minima. Since we are mainly interested in the deep minima which correspond to small values of $\mathcal{M}$ and $\mathcal{N}$, the choice of both $T$ and the initial guess $E_0(t)$ for the gradient descent plays an important role. As we will see below, a poor choice of $E_0(t)$ may not allow finding solutions with particularly low energy $\mathcal{N}$, while poor choices of $T$ may lead to extremely slow convergence.

Consider first an initial condition $E_0(t)$ that is close to a defibrillating signal with a relatively high energy.
As we found previously, a single pulse with $(E^*,t^*)=(0.5\;{\rm V}/{\rm cm}, 20\;{\rm ms})$ does not produce defibrillation, but it can become a defibrillating signal if small adjustments are applied to either $E^*$ or $t^*$. Alternatively, the electric field $E_0(t)$ corresponding to such a single pulse can be modified slightly over a finite time interval by applying adjoint optimization. For instance, after a minimization over the interval $0 \le t \le T$, $T=300\;{\rm ms}$, a defibrillating signal was found with energy $\mathcal{N}= 0.926\;{\rm V}^2\cdot{\rm ms}/{\rm cm}^2$, which corresponds to 38\% of the single-pulse energy $\mathcal{N}_1$ and 1\% of the characteristic LEAP energy $\mathcal{N}_5$. 
For reference, the following hyperparameters were used: $\Delta s = 10^{-5}\;({\rm V}/{\rm cm})^2$, $\alpha = 50\;{\rm cm}^2/({\rm V}^2\cdot {\rm ms})$, and $(\gamma_1,\gamma_2,\gamma_3) = (1,0.2,0.2)$.

\begin{figure} 
  \centering
  \includegraphics[width=\columnwidth]{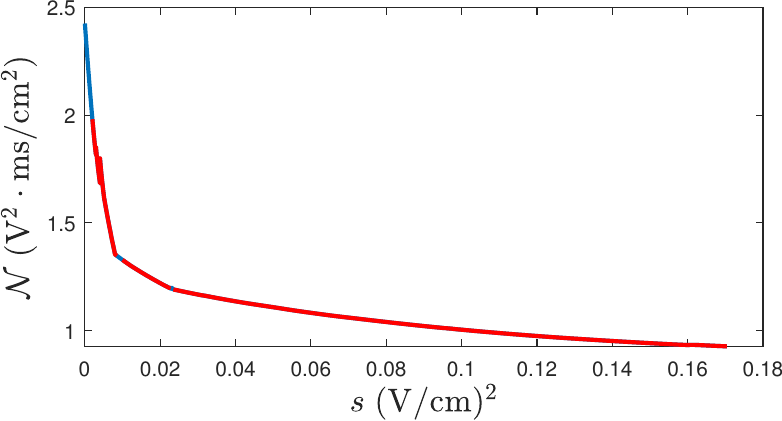}
  \caption{The energy $\mathcal{N}$ as function of $s$ for initial electric field $E_0(t)$ corresponding to the single pulse with $(E^*,t^*)=(0.5\;{\rm V}/{\rm cm}, 20\;{\rm ms})$. The electric field was optimized over an interval with $T=300\;{\rm ms}$. The line color indicates unsuccessful (blue) or successful (red) defibrillation.}
  \label{fig:calN_improved_pulse}
\end{figure}

\begin{figure} 
  \centering
  \includegraphics[width=\columnwidth]{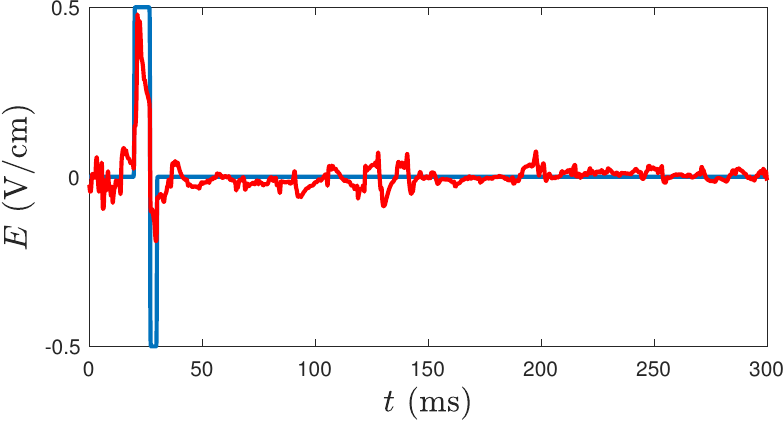}
  \caption{Nondefibrillating single pulse (blue) and defibrillating protocol of lower energy found by adjoint optimization (red).}
  \label{fig:Et_improved_single_pulse}
\end{figure}

During the minimization, both the spatial heterogeneity $\mathcal{M}$ and the energy $\mathcal{N}$ gradually decrease. The latter is shown in Figure \ref{fig:calN_improved_pulse},  with the color-coding describing the result (success/failure of defibrillation). While electric field profiles close to $E_0(t)$ fail to defibrillate the tissue, further minimization identifies electric field profiles that succeed in defibrillation, with the energy gradually decreasing and the likelihood of unsuccessful defibrillation vanishing. (Since a decrease in $\mathcal{L}$ does not guarantee a decrease in $\mathcal{M}$, there is a small likelihood that a decrease in the energy $\mathcal{N}$ may lead to an increase in $\mathcal{M}$ which corresponds to unsuccessful defibrillation over a narrow range of the optimization parameter $s$, as illustrated in Figure \ref{fig:calN_improved_pulse}.) 
The initial and final electric field profiles are compared in Figure \ref{fig:Et_improved_single_pulse}. Note that the decrease in the energy is mostly due to the decrease in the strength of the electric field over the interval corresponding to the single biphasic pulse.

Defibrillating protocols of substantially lower energy can be found using an initial condition with zero energy, i.e., $E_0(t)=0$. We have investigated single-shot adjoint optimization for $T$ varying from 200 ms to 600 ms in increments of 20 ms. In every case, the energy $\mathcal{N}$ initially increased (with a concomitant decrease in $\mathcal{L}$). However, a defibrillating protocol was only obtained for $T=200$ ms, 380 ms, and 400 ms, with the smallest energy achieved for the shortest $T$ considered, illustrating the very rugged shape of the cost functional in the space of continuous functions $E(t)$. 
The energy can be reduced further by extending the range of time over which the optimization is performed, resulting in a multi-stage procedure illustrated below.

Figure \ref{fig:calN_cont} shows $\mathcal{N}$ as a function of $s$ for an optimization procedure with three stages corresponding to $T=200$ ms, followed by $T=220$ ms and, ultimately, $T=240$ ms, with the instances when $T$ is changed marked by the vertical dashed lines. In the first stage, the energy $\mathcal{N}$ initially increases until a successful defibrillation protocol is found at $s=3.2$ (V/cm)$^2$.
The energy then starts to decrease without an increase in $\mathcal{M}$, so all subsequently found electric field profiles lead to defibrillation. After a fast decrement, $\mathcal{N}$ stagnates around $s=4$ (V/cm)$^2$. 
To attain a further reduction of $\mathcal{N}$, $T$ is increased from 200 ms to 220 ms. After a quick drop, the energy starts to decrease quite slowly, prompting a further increase in $T$ to $240$ ms at $s=17.35$ (V/cm)$^2$. This allowed the energy to be reduced even further, with $\mathcal{N}$ eventually stagnating at the value of around 0.22 V$^2\cdot$ms/cm$^2$, an improvement of more than an order of magnitude compared with the energy of the single-pulse protocol.
Note that, in the third stage, the step size had to be reduced by almost two orders of magnitude compared to the second stage, so the number of iterations, and hence the computational cost, of the latter two stages is comparable.
The values of the hyperparameters for the three stages are: $\Delta s = \{10^{-4}, 5\times 10^{-5}, 10^{-6}\}\;({\rm V}/{\rm cm})^2$ and $\alpha=\{1, 5, 40\}\;{\rm cm}^2/({\rm V}^2\cdot{\rm ms})$. For each stage, $(\gamma_1,\gamma_2,\gamma_3)=(1, 0.2, 0.2)$. 
 
The electric field profile at the end of the first stage ($s=4\;({\rm V}/{\rm cm})^2$) and the electric field at the end of the third stage ($s=17.54\;({\rm V}/{\rm cm})^2$) are shown in Figure \ref{fig:Et_200_240ms}(a). Both are defibrillating protocols, which lead to termination of the electrical activity at $t=320\;{\rm ms}$. 
Two features of the latter electric field profile are worth pointing out. First, similar to biphasic defibrillation protocols, it features a pronounced positive ``pulse'' at $t\approx 70$ ms, immediately followed by a smaller negative ``pulse'' as shown in Figure \ref{fig:Et_200_240ms}(b). Second, the electric field essentially vanishes after $t\approx 130$ ms (i.e., roughly one-half of the period, $P$), far sooner than the end of the optimization interval $T=240$ ms. Hence, successful defibrillation can be achieved using low energy {\it and} very quickly.

\begin{figure} 
  \centering
  \includegraphics[width=\columnwidth]{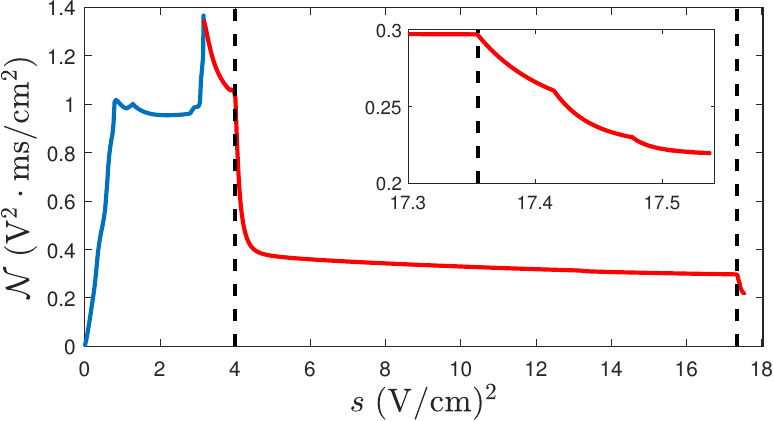} \\ \hspace{2mm}
  \caption{(a) $\mathcal{N}$ as a function of $s$ for an optimization sequence with three stages of different $T$-values, separated by dashed lines. In the first stage, $T=200\;{\rm ms}$ and the starting electric field $E_0(t)=0$, for $0 \le t \le T$. For the second and third stages, $T=220\;{\rm ms}$ and $240\;{\rm ms}$, respectively. The red (blue) line color indicates the electric fields $E_s(t)$ which produce (do not produce) defibrillation. The inset shows the zoomed-in view of the short third stage.} 
  
\label{fig:calN_cont}
\end{figure}
\begin{figure} 
  \centering
  \includegraphics[width=\columnwidth]{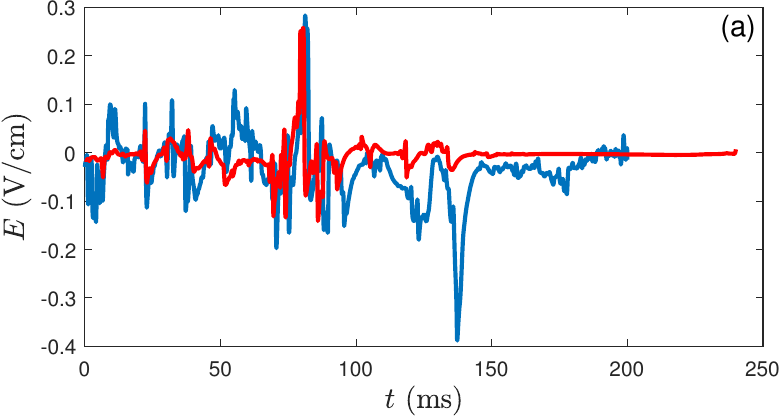} \\ \vspace{3mm}
  \includegraphics[width=\columnwidth]{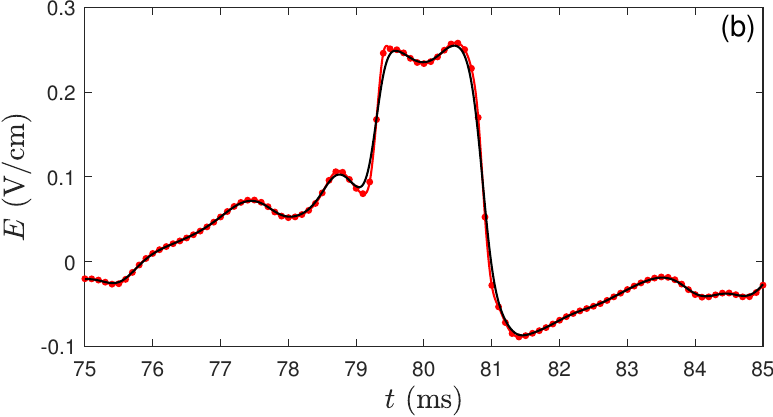}
  \caption{(a) Electric fields found by the gradient descent sequence presented in Figure \ref{fig:calN_cont}, at $s=4\;({\rm V}/{\rm cm})^2$ (blue) and $s=17.54\;({\rm V}/{\rm cm})^2$ (red). (b) The portion of the red signal in panel (a) corresponding to the large spoke. The dots represent the value of $E(t)$ at $t_l=l\Delta t$, $l$ integer.  and the solid line is the cubic spline used to interpolate the value of $E(t)$ at each Runge-Kutta step. A slightly smoothed signal $E'(t)$ (with $\sigma=0.1$ ms) is shown in black.}
  \label{fig:Et_200_240ms}
\end{figure}

\subsection{Improved gradient descent}
\label{sec:improv_grad}

The speed of convergence of the gradient descent iteration \eqref{eq:grad_desc} can be increased significantly using the Nesterov accelerated gradient (NAG) method \cite{liu2022,su2016}. In this work, we used the following implementation of NAG
\begin{align}
    F_{s+\Delta s}(t) &= E_s(t) - \Delta s\, \mathcal{G}(t)|_{E_s(t)},\nonumber\\
  \label{eq:nag}
    E_{s+\Delta s}(t) &= F_{s+\Delta s}(t) + \beta (F_{s+\Delta s}(t)-F_s(t)),
\end{align}
where $F_0(t)=E_0(t)$ and $0\le\beta\le 1$ is a tunable parameter. In particular, $\beta=0$ corresponds to the standard gradient descent algorithm. Smaller values of $\beta$ lead to a slower rate of change of $E_s(t)$, making the method more stable.
Here we illustrate NAG by identifying a defibrillating protocol on a long temporal interval with $T=600\;{\rm ms}$, where standard gradient descent fails. As we will see below, an increase in $T$ allows further substantial reduction in the energy.

\begin{figure} 
  \centering
  \includegraphics[width=\columnwidth]{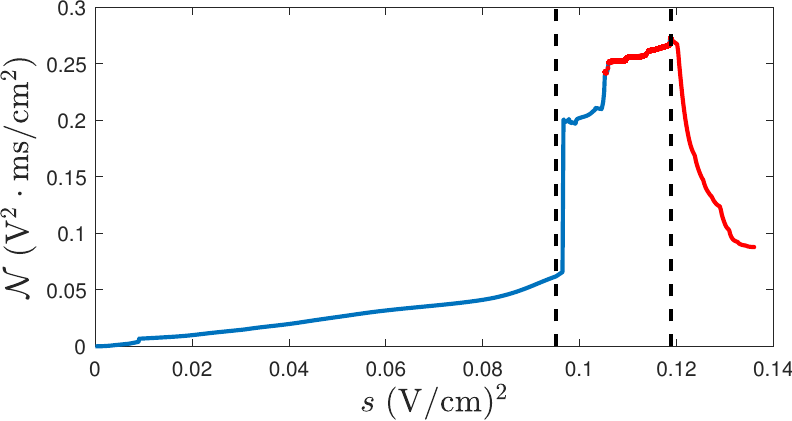}
  \caption{$\mathcal{N}$ as a function of $s$ for an optimization sequence with three stages, separated by dashed lines. In the first stage, the optimization was performed by gradient descent with $T=600\;{\rm ms}$ and $E_0(t)=0$, $0 \le t \le T$. In the second stage, NAG was used with $T$ kept at $600\;{\rm ms}$. The final stage continued to use NAG, but $T$ was extended to $820\;{\rm ms}$. The red (blue) line color indicates the electric fields $E_s(t)$ which produce (do not produce) defibrillation.}
  \label{fig:calN_600ms_cont}
\end{figure}

We again start with $E_0(0)=0$ for $0 \le t \le T$ and apply the standard gradient descent algorithm (with $\alpha=1\;{\rm cm}^2/({\rm V}^2\cdot{\rm ms})$, $(\gamma_1,\gamma_2,\gamma_3)=(1, 0.2, 0.2)$, and $\Delta s=10^{-6}\;({\rm V}/{\rm cm})^2$), which finds electric field profiles with increasing energy that fail to achieve defibrillation, as shown in Figure \ref{fig:calN_600ms_cont}. At $s=0.095\;({\rm V}/{\rm cm})^2$, we switched to NAG with $\beta=0.5$, which quickly identified a family of defibrillating protocols (corresponding to $s\ge0.108\;({\rm V}/{\rm cm})^2$). Note that by simply continuing with the standard gradient descent, no defibrillating electric field was identified up to $s=0.12\;({\rm V}/{\rm cm})^2$. Hence, NAG was essential for finding a defibrillating protocol.

\begin{figure} 
  \centering
  \includegraphics[width=\columnwidth]{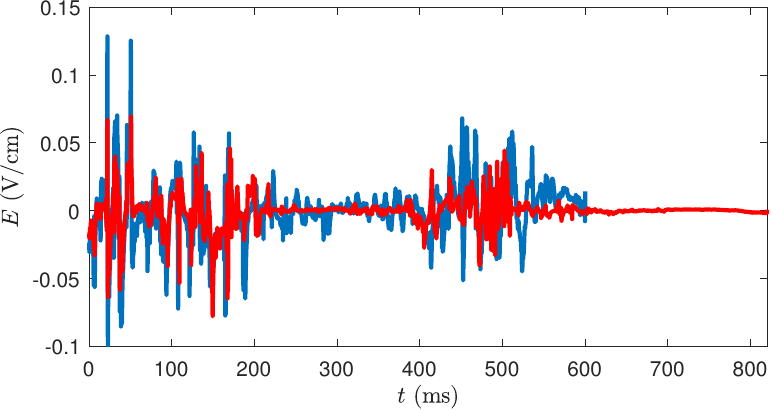}
  \caption{Electric fields $E(t)$ corresponding to $s=0.108\;({\rm V}/{\rm cm})^2$ (blue)  and $s=0.136\;({\rm V}/{\rm cm})^2$ (red) in Figure \ref{fig:calN_600ms_cont}.}
  \label{fig:Et_600ms}
\end{figure}

The electric field profile corresponding to $s=0.108\;({\rm V}/{\rm cm})^2$ is shown in Figure \ref{fig:Et_600ms} (blue line); its energy is $\mathcal{N}=0.24\;{\rm V}^2\cdot{\rm ms}/{\rm cm}^2$, comparable to that of the best protocol shown in Figure \ref{fig:Et_200_240ms}. Further optimization using NAG with the same hyperparameters lead to a decrease in $\mathcal{M}$ at the cost of a slightly increasing $\mathcal{N}$, so we chose to extend $T$ from $600\;{\rm ms}$ to $820\;{\rm ms}$ at $s=0.119\;({\rm V}/{\rm cm})^2$. This resulted in a fast decrease of $\mathcal{N}$, as shown in Figure \ref{fig:calN_600ms_cont}. The final electric field profile (red line in Figure \ref{fig:Et_600ms}) has energy $\mathcal{N}=0.088\;{\rm V}^2\cdot{\rm ms}/{\rm cm}^2$, which corresponds to 3\% of the single-pulse energy $\mathcal{N}_1$ and 0.1\% of the characteristic LEAP energy $\mathcal{N}_5$.

It is worth pointing out that while NAG can substantially increase the rate of convergence, it requires a careful choice of hyperparameters $\beta$ and $\Delta s$ to ensure stability. For instance, NAG iterations starting with the electric field corresponding to $s=0.095\;({\rm V}/{\rm cm})^2$ in Figure \ref{fig:calN_600ms_cont} and $\beta = 0.9$ lead to such an unstable behavior. It remains to be explored whether there are adaptive implementations of NAG with variable $\beta$ and $\Delta s$ that are both faster than standard gradient descent and stable.

\section{Discussion and conclusions}
\label{sec:conclusions}

In this study, a simple two-dimensional numerical model of atrial tissue containing anatomical heterogeneities -- the essential ingredient responsible for the emergence of virtual electrodes -- was used to explore ultra-low-energy defibrillation. The approach considered here is impractical -- it requires an accurate mathematical model of the tissue as well as complete knowledge of the state of the tissue at the initial time. Furthermore, computation of a defibrillating electrical field cannot be performed in real time. However, our results provide a number of important lessons.

Despite the state of fibrillation -- in the model and for the parameters considered here -- persisting over a very long temporal interval, fibrillation appears to be a long chaotic transient rather than a sustained state. This implies that fibrillation will eventually disappear on its own, so the lowest defibrillation energy is zero. Indeed, we found that, using adjoint optimization, the energy required for defibrillation can be reduced by three(!) orders of magnitude compared with current state-of-the-art protocols such as LEAP. There is little doubt in that this energy could be reduced even further by using better initial conditions, improved gradient descent algorithms, or additional tuning of the hyperparameters (e.g., the choice of the weights $\alpha$ and $\gamma_i$ used in defining the cost function).

All of the defibrillating signals identified via adjoint optimization feature high-frequency oscillation with time scale of order 10 ms. Note that this time scale is essentially identical to the duration of an optimal biphasic pulse \cite{bragard2013shock}.
This time scale is also comparable to that describing the dynamics of the fast gating variable ($u_2$) which controls the repolarization dynamics and conduction velocity. In contrast, the slow gating variable ($u_3$) which controls the action potential duration evolves on a much longer time scale. Hence, the applied electric field largely plays the role of the fast gating variable in modulating the slow outward current. These fast oscillations appear to be an essential property of defibrillating protocols rather than an artifact of adjoint optimization: applying even a minor amount of smoothing 
\begin{align}
    E'(t)=\frac{1}{\sqrt{\pi}\sigma}\int_{-\infty}^\infty e^{-(t-t')^2/\sigma^2}E(t')dt'
\end{align}
to the computed electric field, with $\sigma$ of order 0.1 ms,
leads to a failure of defibrillation. 

Given that the maximal strength of the electric field is quite small, it is clear that virtual electrodes do not generate new wavefronts and phase singularities. Rather, the effect of the electric field is to gently alter the motion of existing phase singularities. Since phase singularities cannot be created or destroyed by weak electric fields, defibrillation can be achieved by one or both of the following mechanisms: (i) individual phase singularities, regardless of their topological charge, can be swept across the boundary, i.e., outside of the tissue, or (ii) pairs of phase singularities with opposite topological charges can be mutually annihilated inside the tissue. While drift and interaction of spiral waves induced by near-resonant time-periodic electric fields are well-understood \cite{biktashev1994,biktashev1995}, the impact of aperiodic high-frequency electric fields has received little attention so far. 

\begin{figure*} 
  \centering
  \subfigure[]{\includegraphics[width=5.7cm]{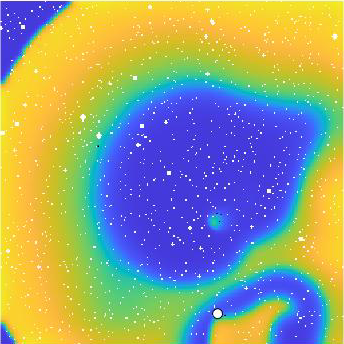}}
  \hspace{1mm}
  \subfigure[]{\includegraphics[width=5.7cm]{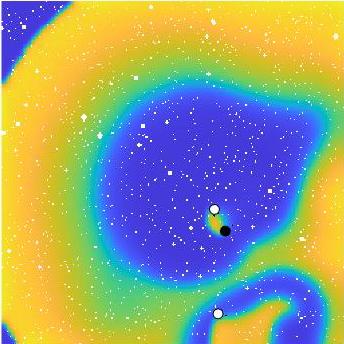}}
  \hspace{1mm}
  \subfigure[]{\includegraphics[width=5.7cm]{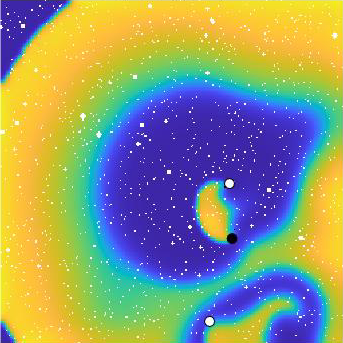}}
 \caption{(Multimedia available online) Snapshots of the voltage field ($u_1$) at $t=174$ ms for three different choices of the applied electric field: (a) electric field profile $E(t)$ shown in red in Figure \ref{fig:Et_200_240ms}, (b) its smoothed version $E'(t)$ with $\sigma=0.1$ ms, (c) no electric field, i.e., $E(t)=0$. The white and black circles mark the phase singularities with positive (negative) topological charge. The same colorbar as in Figure \ref{fig:snap} is used. The remaining phase singularity at the bottom of the panels will leave the domain by $t=203$ ms in every case.}
  \label{fig:3cases}
\end{figure*}

This study provides the first insight into how weak aperiodic electric fields can affect the dynamics. The movies illustrating the evolution of the excitation waves with the defibrillating electric field $E(t)$ shown in red in Figure \ref{fig:Et_200_240ms}, with its slightly smoothed version $E'(t)$, as well without an applied electric field are included as multimedia files (available online). These movies also show the dynamics of the associated phase singularities computed using the level-set-based approach \cite{gurevich2019}. These movies clearly demonstrate that defibrillation is achieved through mechanism (ii). Figure \ref{fig:3cases} shows the snapshots of the state for the three cases about 30 ms after
a short wavefront (connecting a pair of nearby phase singularities with opposite topological charge, shown as white/black circles) passes through a region of nearly refractory tissue behind the trailing edge of another excitation wave.

This region is characterized by low conduction velocity and low excitability, so any wave passing through it is extremely susceptible to very small perturbations. Such vulnerable windows \cite{starmer1993} have been a target of defibrillation protocols for a long time \cite{starmer1997,biktashev1998,yamanouchi2001}. 
Without an applied electric field, the excitation wave manages to pass through with the two phase singularities never approaching each other too closely. The defibrillating signal $E(t)$ delays repolarization, slowing down the excitation wave sufficiently to produce conduction block and annihilation of the two phase singularities. The smoothed signal $E'(t)$ brings the two phase singularities closer to each other but not enough to lead to their annihilation. The conditions under which phase singularities merge and an excitation wave collapses were discussed in Ref. \onlinecite{marcotte2017}.

The possibility of leveraging this defibrillation mechanism raises an interesting question regarding the statement of the optimization problem. Fibrillation can be terminated by decreasing spatial gradients of the voltage and the gating variables, as shown here, but the same result can also be achieved by eliminating phase singularities. The latter objective can be targeted directly by a suitable modification of the cost function. For instance, a positive-definite function $\mathcal{M}=|\nabla u_i\times\nabla u_j|^2$ with $i\ne j$ will be strongly peaked at the cores of the spiral waves surrounding phase singularities. Minimization of a respective cost function should produce electric fields that lead to pair-wise elimination of counter-rotating spiral waves. This can produce even more efficient defibrillation protocols and is worth being explored further.

\begin{acknowledgments}
We would like to thank Christopher Marcotte for bringing to our attention a WebGL solver for the Fenton-Karma model \cite{webgl2024fentonkarma}, which was used for exploring the behavior of solutions for different parameter sets. We would also like to thank Daniel Gurevich for help with the topological analysis of the data.

\end{acknowledgments}

\section*{Author declarations}
\subsection*{Conflict of interest}
The authors have no conflicts to disclose.

\subsection*{Author contributions}
{\bf A. Garz\'on:} Conceptualization (equal); Investigation; Methodology (lead); Software; Visualization (lead); Writing/Original Draft Preparation (lead); Writing/Review \& Editing (equal). 
 {\bf R. Grigoriev:} Conceptualization (equal); Formal Analysis, Methodology (supporting); Resources; Visualization (supporting); Writing/Original Draft Preparation (supporting); Writing/Review \& Editing (equal).

\section*{Data Availability Statement}

The data that support the findings of this study are available from the corresponding author upon reasonable request.

\appendix
\section{Computation of the functional derivative}
\label{sec:fin_diff}

Note that the cost function $\mathcal{L}$ depends on $E(t)$ both explicitly through $\mathcal{N}[E]$ and implicitly through $\mathcal{M}[{\bf u}]$, so its functional derivative $\mathcal{G}=\delta\mathcal{L}/\delta E$ cannot be written in an explicit analytical form as is often the case in variational calculus. 
Instead we can compute it using the formal definition \cite{parr1989}
\begin{align}
  \label{eq:fun_der}
  \left\langle \mathcal{G}, \phi \right\rangle = \left(\frac{d}{d\varepsilon} \mathcal{L}\left[E(t) + \varepsilon \phi(t)\right]\right)_{\varepsilon = 0},
\end{align}
where $\phi(t)$ is an arbitrary function and, in this section, the angle brackets denote the Euclidean inner product
\begin{align}
  \label{eq:scalar_inner}
  \left\langle f, g \right\rangle = \int_0^T f(t)g(t)\,dt.
\end{align}
For a numerical solution evaluated at a discrete set of time steps $t_l=l\Delta t$ with $l=0,\cdots,N$, we can evaluate the function derivative by using $\phi_l(t) = \delta(t-t_l)$, where $\delta(t)$ is the Dirac delta function, such that
\begin{align*}
  \left\langle \mathcal{G}, \phi_l \right\rangle
  = \int_0^T \mathcal{G}(t)\, \delta(t-t_l)\,dt 
  = \mathcal{G}(t_l).
\end{align*}
Evaluating the derivative in the definition \eqref{eq:fun_der} using finite differences, we find
 \begin{align}
   \label{eq:fin_diff}
    \mathcal{G}(t_l) \approx \frac{\mathcal{L}[E(t) + \varepsilon{\delta}(t-t_l)] - \mathcal{L}[E(t)]}{\varepsilon}
  \end{align}
for a sufficiently small $\varepsilon$.
Note that in order to evaluate the functional derivative using \eqref{eq:fin_diff} at any one time step, the solution ${\bf u}(t,{\bf r})$ should be evaluated at all $N$ time steps. Hence the total number of operations involved in evaluating $\mathcal{G}(t)$ on the interval $t\in [0,T]$ scales as $N(N+1)\sim N^2$.

\section{Spatial discretization}
\label{sec:discretization}

For the purpose of spatial discretization, the circular conduction heterogeneities were approximated by polygons with sides parallel to either the $x$ or $y$ axis. The corresponding normal vector ${\bf n}$ is therefore equal to $\pm{\bf\hat{y}}$ or $\pm{\bf\hat{x}}$, respectively.

For interior points, the Laplacian $\nabla^2 u_i$ was approximated using the second-order centered finite differences formula
\begin{align}
  \nabla^2 u_i({\bf r}_j) \approx \frac{u_{i,j_L}+ u_{i,j_T}- 4u_{i,j}+ u_{i,j_B} + u_{i,j_R}}{\left(\Delta x\right)^2},
  \label{eq:lap_fin_diff}
\end{align}
where $j_L,j_T,j_B,$ and $j_R$ are respectively the indices of the left, top, bottom, and right neighbor of the $j$-th grid point. For points on the boundary, the missing neighbor(s) are replaced by fictitious ones and the boundary conditions \eqref{eq:nu1E} and \eqref{eq:nui} are used to compute the value of $u_i$ at those ghost points \cite{Thomas1995}.   For instance, if the point $j$ is on a boundary segment with ${\bf n} = {\bf\hat{x}}$, its right neighbor is a ghost point. Assuming no other neighbor is missing, the Laplacian $\nabla^2 u_1$ is approximated by the analogue of Equation \eqref{eq:lap_fin_diff},
\begin{align}
  \nabla^2 u_1({\bf r}_j) \approx \frac{u_{1,j_L}+ u_{1,j_T}- 4u_{i,j}+ u_{i,j_B} + u_1^G}{\left(\Delta x\right)^2},
  \label{eq:lap_bound}
\end{align}
where $u_1^G$ is the value of $u_1$ at the ghost right neighbor. To compute $u_1^G$, notice that, when ${\bf n} = {\bf\hat{x}}$,  the boundary condition \eqref{eq:nu1E} reduces to
\begin{align}
  \label{eq:pupx_E}
  \frac{\partial u_1}{\partial x} - E = 0.
\end{align}
Approximating the derivative in \eqref{eq:pupx_E} with the second-order centered finite differences formula, we get
\begin{align}
  \frac{u_1^G-u_{1,j_L}}{2\Delta x} - E \approx 0.
  \label{eq:bound_ghost}
\end{align}
Solving for $u_1^G$ in \eqref{eq:bound_ghost} and substituting the result into \eqref{eq:lap_bound}, we have
\begin{align}
  D_1 \nabla^2 u_1({\bf r}_j) \approx \sigma \left(2u_{1,j_L} + u_{1,j_T} -4 u_{1,j} + u_{1,j_B}\right) + \mu E,
  \label{eq:D1nabla}
\end{align}
where
\begin{align}
\sigma=\frac{D_1}{(\Delta x)^2}~~{\rm and}~~\mu = \frac{2D_1}{\Delta x}.
\end{align}
Grid points missing other neighbors are dealt with analogously. The discretization of $\nabla^2u_i$, $i \ge 2$,  for boundary points, where \eqref{eq:nui} holds, produces expressions similar to \eqref{eq:D1nabla} but without the term involving the electric field. Consequently, the discretization of the term $\wt{L}{\bf u}$ in Equation \eqref{eq:puwtLu} can be written in the form
\begin{align}
  \label{eq:Lbfu}
  L{\bf w} + E{\bf b},
\end{align}
where $L$ is an $mn \times mn$ block-diagonal matrix (with diagonal blocks of size $m\times m$) and ${\bf b}$ is a column vector with $mn$ elements. Hence, the evolution equation takes the form
\begin{align}
  \label{eq:dot_bfu}
  \dot{\bf w} \approx L{\bf w} + F({\bf w}) + E{\bf b},
\end{align}
where the column vector $F$ describes the ionic model.

One can similarly write the functionals $\mathcal{J}_i$ in terms of the discretized state vector. Let ${\bf w}_i = [u_{i,1}, u_{i,2}, ..., u_{i,m}]^\tp$ such that ${\bf w} =[{\bf w}_1^\tp, {\bf w}_2^\tp, ..., {\bf w}_n^\tp]^\tp$. 
Furthermore, let us denote the values of the spatial derivatives $\partial_xu_i$ and $\partial_yu_i$ on the computational grid,  respectively, as $G_x {\bf w}_i$ and $G_y {\bf w}_i$, where $G_x$ and $G_y$ are $m \times m$ matrices representing the second-order finite differences approximations of the derivatives. With the help of this notation, we can write
\begin{align*}
  \mathcal{J}_i 
   \approx \left[{\bf w}_i^\tp \,S\,  {\bf w}_i\right]_{t=T},
\end{align*}
where $S=\left[G_x^\tp G_x + G_y^\tp G_y\right](\Delta x)^2$, and 
\begin{align}
  \label{eq:calL0}
  \mathcal{L} 
  \approx \frac{1}{2} \left[{\bf w}^\tp R {\bf w}\right]_{t=T} + \frac{\alpha}{2} \int_0^T E^2 dt,
\end{align}
where $R$ is an $nm \times nm$ block-diagonal matrix with the $i$-the $m \times m$ diagonal block equal to $\gamma_i S$, $i=1,2,...,n$.

\section{Derivation of the adjoint equations}
\label{sec:adj_method}

In this section, we consider a minimization problem for the functional \eqref{eq:calL0} subject to the dynamical constraint \eqref{eq:dot_bfu} for all $t\in[0,T]$ which we will rewrite as
\begin{align}
  \label{eq:wEv}
  {\bf q}[E,{\bf w}] = \dot{\bf w} - L{\bf w} - F({\bf w}) - E{\bf b} = {\bf 0}.
\end{align}
By introducing a Lagrange multiplier $\bm{\lambda}(t)$ it can be converted to an unconstrained minimization problem for the new functional 
\begin{align}
  \label{eq:calL3}
  \mathcal{L}'[E,{\bf w},\bm{\lambda}] = \mathcal{L} -
  \left\langle \bm{\lambda},{\bf q}\right\rangle,
\end{align}
where, in this section, the angle brackets denote the inner product
\begin{align}
  \left\langle{\bf f},{\bf g}\right\rangle = \int_0^T [{\bf f}(t)]^{\tp}{\bf g}(t)\,dt.
\end{align}
Note that, for ${\bf w}$ which satisfies the governing equations, ${\bf q}=0$, so $\mathcal{L}'=\mathcal{L}$. 
Applying the chain rule, we find
\begin{align}
\label{eq:crule}
    \frac{\delta\mathcal{L}}{\delta E(t)} = \frac{\partial \mathcal{L}'}{\partial E(t)} 
    &+\int_0^T\frac{\partial\mathcal{L}'}{\partial {\bf w}(s)} \frac{\delta{\bf w}(s)}{\delta E(t)} ds \nonumber\\
    &+\int_0^T\frac{\partial\mathcal{L}'}{\partial \bm{\lambda}(s)} \frac{\delta\bm{\lambda}(s)}{\delta E(t)} ds,
\end{align}
where the last term on the right-hand side vanishes, since 
\begin{align}
    \frac{\partial\mathcal{L}'}{\partial \bm{\lambda}} = {\bf q} = {\bf 0}.
\end{align}
The second term can also be eliminated by requiring
\begin{align}
  \label{eq:null}
  \frac{\partial\mathcal{L}'}{\partial{\bf w}} = {\bf 0}.
\end{align}

Holding variables ${\bf w}$ and $\bm{\lambda}$, 
the functional derivative can now be evaluated using the definition \eqref{eq:fun_der}:
\begin{align}
  \label{eq:dL0dE_alphaE}
    \mathcal{G}=\frac{\delta\mathcal{L}}{\delta E}=
    \frac{\partial\mathcal{L}'}{\partial E} = \alpha E + \bm{\lambda}^\tp{\bf b},
\end{align}
which requires the knowledge of $\bm{\lambda}(t)$ on the interval $t\in[0,T]$. The governing equations for the Lagrange multiplier can be derived using the condition \eqref{eq:null}:
\begin{align}
  \label{eq:lin}
  \frac{1}{2}\frac{\partial}{\partial{\bf w}} \left[{\bf w}^\tp R {\bf w}\right]_{t=T} &-
  \frac{\partial}{\partial{\bf w}} \left\langle \bm{\lambda}, \dot{\bf w}\right\rangle\nonumber\\ 
  &+\frac{\partial}{\partial{\bf w}} \left\langle \bm{\lambda}, L{\bf w} + F({\bf w})\right\rangle =
  {\bf 0},
\end{align}
where the temporal derivative can be moved from ${\bf w}$ to $\bm{\lambda}$ via integration by parts:
\begin{align}
  \label{eq:lamb_dot_v}
  \left\langle\bm{\lambda},\dot{\bf w}\right\rangle =
  \bm{\lambda}^\tp(T){\bf w}(T) -
  \bm{\lambda}^\tp(0){\bf w}(0) -
  \left\langle \dot{\bm{\lambda}}, {\bf w} \right\rangle.
\end{align}
It is straightforward to show that
\begin{subequations}
  \label{eq:derivs}
\begin{align}
  \frac{\partial}{\partial{\bf w}(T)}\left[{\bf w}^\tp R {\bf w}\right]_{t=T} &= 2R{\bf w}(T),\\
  \frac{\partial}{\partial{\bf w}(T)} \left[ \bm{\lambda}^\tp(T){\bf w}(T) \right] &= \bm{\lambda}(T),\\
  \label{eq:plampv0}
  \frac{\partial}{\partial{\bf w}(0)} \left[ \bm{\lambda}^\tp(0){\bf w}(0) \right] &= {\bf 0},\\
  \frac{\partial}{\partial{\bf w}(t)} \left\langle \dot{\bm{\lambda}}(t), {\bf w}(t) \right\rangle &= \dot{\bm{\lambda}}(t),\\
  \frac{\partial}{\partial{\bf w}(t)} \left\langle \bm{\lambda}(t), L{\bf w}(t) + F({\bf w}(t))\right\rangle &=\nonumber\\ L^\tp\bm{\lambda}(t) &+
 J_F^\tp \bm{\lambda}(t),
\end{align}
\end{subequations}
where Equation \eqref{eq:plampv0} is a consequence of the 
fixed initial condition ${\bf w}(0)={\bf w}_0$. 
Evaluating the partial derivatives in Equation \eqref{eq:lin} at $t=T$ yields the initial condition \eqref{eq:main_lambT}, while evaluating the partial derivatives for $0<t<T$ yields the evolution equation \eqref{eq:main_dot_lamb}.

Let us conclude this section by estimating the number of operations required to evaluate the functional derivative $\mathcal{G}$ using the adjoint method. Integration of the evolution equation \eqref{eq:main_dot_bfu} to compute ${\bf w}(t)$ on the interval $[0,T]$ requires $O(N)$ operations. Integration of Equation \eqref{eq:main_dot_lamb} to compute $\bm{\lambda}(t)$  on the interval $[0,T]$ also requires $O(N)$ operations, although, in the latter case, one time step is more costly than in the former case due to the larger number of arithmetic and memory transfer operations required to calculate the $n^2$ elements of the Jacobian $J_F$ compared with just $n$ elements of $F$. Nonetheless, since the total number of operations scales linearly with $N$, the adjoint method is substantially less expensive than the finite differences method described in Appendix \ref{sec:fin_diff} which requires a number of operations that scales as $N^2$.

\section{Computation of the spectral density}
\label{sec:spectral}
The spectral density $W(f)$ is computed in the following manner. The Fast Fourier Transform $U(f_l,{\bf r}_i)$ of $u_1(t_l,{\bf r}_i)$, where $t_l = l\Delta t,~l=1,2,...,N$, $N=2.5\times 10^5$, and the frequencies $f_l=l/(N\Delta t)$, was computed for a hundred evenly spaced locations ${\bf r}_i$. Then, the local power spectral density $W(f, {\bf r}_i)$ (subscript $l$ dropped) was calculated as $|U(f,{\bf r}_i)|^2$ normalized by the mean squared amplitude of $u_1(t,{\bf r}_i)$ \cite{press1992}. Finally, the global power spectral density $W(f)$, shown in Figure \ref{fig:fourier}, was determined as the average of the $W(f,{\bf r}_i)$ over all the sampled locations ${\bf r}_i$.

\bibliography{references,cardiac}
\end{document}